# Investigation of relaxation phenomena in high-temperature superconductors $HoBa_2Cu_3O_{7-\delta}$ at the action of pulsed magnetic fields

J.G. Chigvinadze*, J.V. Acrivos**, S.M. Ashimov*, A.A. Iashvili*, T. V. Machaidze*, Th. Wolf***

*E. Andronikashvili Institute of Physics, 0177 Tbilisi, Georgia*

**San Jose' State University, San Jose' CA 95192-0101, USA**

***Forschungszentrum Karlsruhe, Institut für Festkörperphysik, 76021 Karlsruhe, Germany***

**Summary**

It is used the mechanical method of Abrikosov vortex stimulated dynamics investigation in superconductors. With its help it was studied relaxation phenomena in vortex matter of high-temperature superconductors. It established that pulsed magnetic fields change the course of relaxation processes taking place in vortex matter. The study of the influence of magnetic pulses differing by their durations and amplitudes on vortex system of isotropic high-temperature superconductors system $HoBa_2Cu_3O_{7-\delta}$ showed the presence of threshold phenomena. The small duration pulses doesn't change the course of relaxation processes taking place in vortex matter. When the duration of pulses exceeds some critical value (threshold), then their influence change the course of relaxation process which is revealed by stepwise change of relaxing mechanical moment $\tau^{rel}$.

These investigations showed that the time for formatting of Abrikosov vortex lattice in $HoBa_2Cu_3O_{7-\delta}$ is of the order of 20 μs which on the order of value exceeds the time necessary for formation of a single vortex observed in type II superconductors.

## 1. Introduction

The present communication is devoted to the experimental investigation of relaxation phenomena in high-temperature superconductors of $HoBa_2Cu_3O_{7-\delta}$ system.

High-temperature superconductors are characterized by such high critical transition temperatures $T_c$ in the superconducting state, they remain superconductors at temperatures when their thermal fluctuations energy becomes compared with the elastic energy, and also with the pinning energy [1]. It creates prerequisites for phase transitions. Due to the layered crystal structure and anisotropy, which is a characteristic high-temperature superconductors, they reveal conditions for the appearance of different phases on *B-T* diagram.( *B* is magnetic induction, *T*-is temperature)[2-13]. As example, Abrikosov vortex lattice begin melting near the critical $T_c$ temperature what is followed by the essential change of vortex continuum flow dynamics along with sharp change of character (dynamics) of relaxation phenomena. In high-temperature superconductors it is observed such relaxation processes as a slow logarithmic decrease of captured flux with time at temperatures much below their superconductive critical transition temperature $T_c$ [14-16]. The logarithmic character of relaxation is explained by the Anderson [17]. Near $T_c$, in the range of Abrikosov vortex lattice melting, the logarithmic character of relaxation is changed by the power one with 2/3 exponent [18].

Consequently, the study of relaxation processes in high-temperature superconductors is an important problem.

## 2. Experimental

For Investigation it was used currentless mechanical method of Abrikosov vortex stimulated dynamics study by magnetic pulses revealing relaxation phenomena in vortex matter described in work [19]. This method is a development of currentless mechanical method of pinning investigations [20,21] and is based on pinning forces countermoments measurements and viscous friction, acting on a axially symmetrical superconducting sample in an outer (transverse) magnetic field. Countermoments of pinning forces and of viscous friction, acting on a superconductive sample from quantized vortex lines side (Abrikosov vortices) are defined the way as it was described in work [22,23]. The sensitivity of the method accordingly works [24], is equivalent to $10^{-8}$ V×cm$^{-1}$ in the method of *V-A* characteristics.

The high-temperature superconducting samples of $HoBa_2Cu_3O_{7-\delta}$ system were prepared by the standard solid state reaction method. Samples were made cylindrical with height L=13mm and diameter d=6mm. Their critical temperature was $T_c$=92 K. The investigated samples were isotropic what was established by mechanical moment $\tau$ measurements appearing $H > H_{c1}$ with the penetration of Abrikosov vortices into a freely suspended on a thin elastic thread superconducting sample. The appearance of such moment $\tau = MH\sin\alpha$, characteristic for anisotropic superconductors, is related with penetrating Abrikosov vortices and the mean magnetic moment $\vec{M}$ of a sample which could deviate on angle $\alpha$ from the direction of outer magnetic field $\vec{H}$. In superconducting anisotropic samples it is presented energetically favorable directions for the arrangement of emerging (penetrating) vortex lines which in their turn are fastened by pinning centers creating aforementioned moment $\tau$. The lack of $\tau$ moment is characteristic for isotropic and investigated by us samples, no matter magnetic field value and its previous orientation in respect to $\vec{H}$ in the axial symmetry plane. Pulsed magnetic fields were created by Helmholtz coils. The value of pulsed magnetic fields was changed in $\Delta h = 2 \div 200 Oe$ limits.

In experiments it was used both single and continuous pulsed with repetition frequency ν from 2.5 s$^{-1}$ to 500s$^{-1}$. The duration Δx of pulses was changed from 0,5 до 500 μs. Magnetic pulse could be directed both parallel (*Δh//H*) and perpendicularly (*Δh⊥H*) to applied steady magnetic field $\vec{H}$, creating mixed state of superconducting sample. The standard pulsed generator and amplifier were used to feed Helmholtz coils. The current strength in coils reached up to 40÷50 A.

Samples were high-temperature superconductors of HoBa$_2$Cu$_3$O$_{7-\delta}$ system placed in the center between Helmholtz coils.

The principal set-up of experiment is shown in fig.1 [19,20]. In experiments it is measured the rotation angle $\varphi_2$ of sample depending on the angle of rotation of a torsion head $\varphi_1$, transmitting the rotation to a sample by means of suspension having the torsion stiffness $K \approx 4 \cdot 10^{-1}$ [dyn•cm], which can be replaced when necessary by a less stiff or stiffer one.

The measurements were carried out at a constant speed of rotation of the torsion head, making ω$_1$=1,8·10$^{-2}$ rad/s. Angles of rotation поворота $\varphi_2$ and $\varphi_1$ were determined with an accuracy of ±4,6·10$^{-3}$ and ±2,3·10$^{-3}$ rad, respectively. The uniformity of the magnetic field's strength along a sample was below $\Delta H / H = 10^{-3}$.

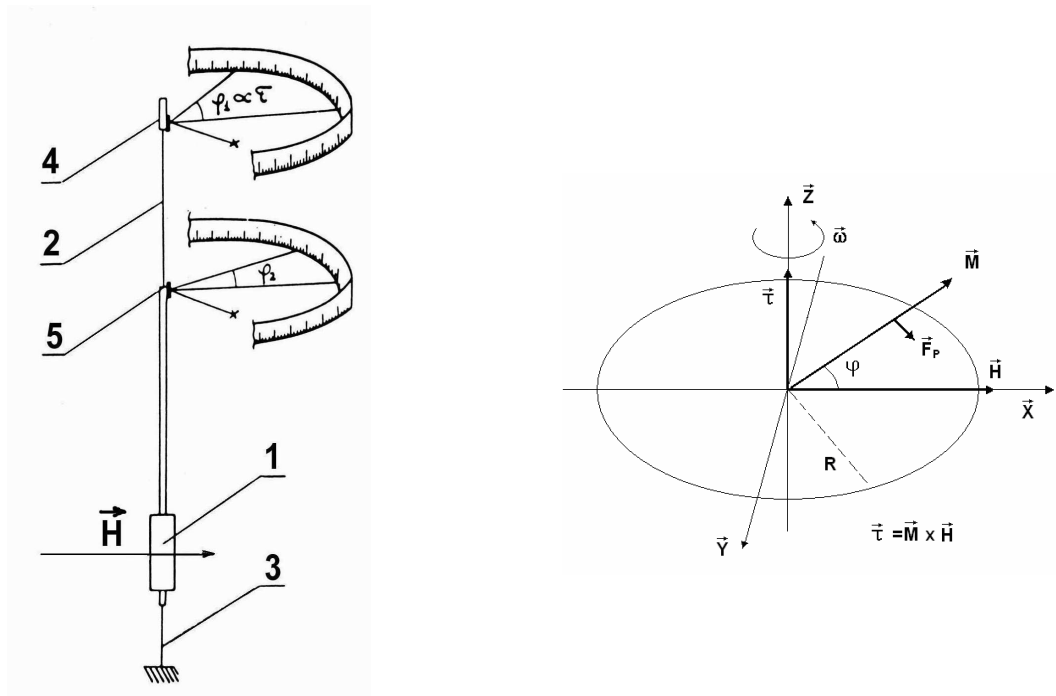

Fig. 1. The schematic diagram and the geometry of the experiment. 1-sample, 2-upper elastic filament, 3-lower filament, 4 - leading head, 5 - glass road. *φ* is angle between $\vec{M}$ and $\vec{H}$

To avoid effects, connected with the frozen magnetic fluxes, the lower part of the cryostat with the sample was put into a special cylindrical Permalloy screen, reducing the Earth magnetic field by the factor of 1200. After a sample was cooled by liquid nitrogen to the superconducting state, the screen was removed, a magnetic field of necessary intensity $H$ was applied and the $\varphi_2(\varphi_1)$ dependences were measured. To carry out measurements at different values of $H$, the sample was brought to the normal state by heating it to до $T > T_c$ at $H = 0$, and only after returning sample and torsion head to the initial state $\varphi_1 = \varphi_2 = 0$, the experiment was repeated.

**3. Results and discussions**

During rotation of the sample both of normal and superconducting states in the absence external magnetic field ($H=0$) the $\varphi_2$ dependence versus $\varphi_1$ is linear and the condition is satisfied.

$$\varphi_1 = \varphi_2 = \omega t$$

The character of the $\varphi_2(\varphi_1)$ dependence is changed significantly, when the sample is in magnetic fields $H > H_{c1}$ at $T < T_c$. Typical $\varphi_2(\varphi_1)$ dependences at T=77K and various magnetic fields for HoBa$_2$Cu$_3$O$_{7-\delta}$ sample ( length of a cylindrical sample L=13mm and diameter d=6mm ) is shown in Fig.2.

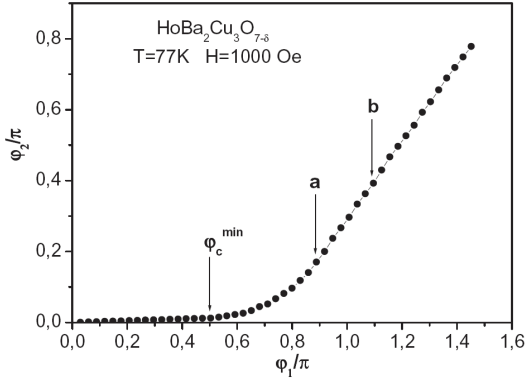

Fig.2. Dependence of the rotation angle of the sample HoBa$_2$Cu$_3$O$_{7-\delta}$ $\varphi_2$ on the rotation angle of the leading head $\varphi_1$ in magnetic field H=1000 Oe at T=77K.

Three distinct regions are observed in Fig.2. In the first (initial) region, the sample does not respond to the increase in $\varphi_1$, i.e. to the applied and increase with time torsion torque as $\varphi_1 \sim \tau = K(\varphi_1 - \varphi_2)$ or responds weakly. Such behavior of the sample can be explained by fact that Abrikosov vortices are not detached from pinning centers at small values of $\varphi_1 \sim \tau$, but if the sample is still turned slightly, this can be caused by elastic deformation of magnetic force lines beyond it or, possibly, by separation of the most weakly fixed vortices. As it is seen from fig.2 , as soon as a certain critical value $\varphi_c^{min}$ depending on $H$ is reached , the first region under goes a transition to the second region in which the velocity of the sample increases gradually with $\varphi_1$ increasing resulting from the progressive process of detachment of vortices from their corresponding pinning centers. One should expect that just in this region, in the rotating sample "the vortices fan" begins to unfold, in with the vortices are distributed according to the instantaneous angles of orientations with respect to the fixed external magnetic field.  In this case the of orientation angles of separate vortex filaments are limited from $\varphi_{fr}$ to $\varphi_{fr} + \varphi_{pin}$, where $\varphi_{fr}$ is the angle on which the vortex filament can be turned with respect to $\vec{H}$ by forces of viscous friction with the matrix of superconductor, and $\varphi_{pin}$ is the angle on with the vortex filament can be turned by the most strong pinning center, studied for the first time in [25].

The gradual transition (at high $\varphi_1$ values) to the third region where the linear $\varphi_2(\varphi_1)$ dependence was observed, allows one to define the countermoments of pinning forces $\tau_p$ and $\tau_{fr}$, independently. Just in this region, when $\omega_1 = \omega_2$ the torque $\tau$, appeared to the uniformly rotating sample, is balanced by the countermoment $\tau_p$ and $\tau_{fr}$. In particular, in the case of continuously rotating sample with frequency $\omega_1 = \omega_2$ one could find similarly to [26,27] the expression for the total braking torque $\tau$ [19] .

Indeed, if we consider in this case a vortex element $d\vec{s}$ moving with velocity $\vec{v}_\perp$ perpendicular to $d\vec{s}$, then the average force acting on this elements is

$$d\vec{f}_v = \vec{v}_\perp \eta ds + \frac{\vec{v}_\perp}{|v_\perp|} F_l ds$$

and the associated braking torque, exerted on the rotating specimen becomes:

$$d\vec{\tau} = \vec{r} \times d\vec{f}_v$$

where $\vec{r}$ is the vector pointing from the rotational axis to the vortex elements, $F_l$ is the pinning force per flux thread per unit length, and $\eta$ is the viscosity coefficient. For a cylindrical specimen of radius $R$ and height $L$ integrating over the individual contribution of all vortex gives a total braking torque $\tau$

$$\tau = \tau_p + \tau_0 \omega \qquad (1)$$

with

$$\tau_p = \frac{4}{3} \frac{BF_l}{\Phi_0} LR^3 ,$$

and

$$\tau_0 = \frac{\pi}{4} \frac{B}{\Phi_0} \eta LR^4 ,$$

Where $B$ is the inductivity averaged over the sample, $\Phi_0$ is the flux quantum , $L$ is the height and $R$ is the radius of the sample.

As it is shown in Fig.2, starting with the point (**a**), where $\omega_1 = \omega_2$, to the superconducting sample uniformly rotating in the homogeneous stationary magnetic field  H=1000 Oe, is applied stationary dynamic torsion moment $\tau_p^{dyn} = \tau_p^{st} + \tau_{fr}$.

If in this region the torsion head is stopped, then at the expense of relaxation processes connected with the presence of viscous forces acting on vortex filaments, the sample will continue the rotation in the same direction (with decreasing velocity)    until it reaches a certain equilibrium position, depending on the $H$ value. The Fig.3 shows curves of $\Delta\varphi_2^{rel}$ time dependences at the stopped leading head for $HoBa_2Cu_3O_{7-\delta}$ sample at T=77K and H=1000 Oe.

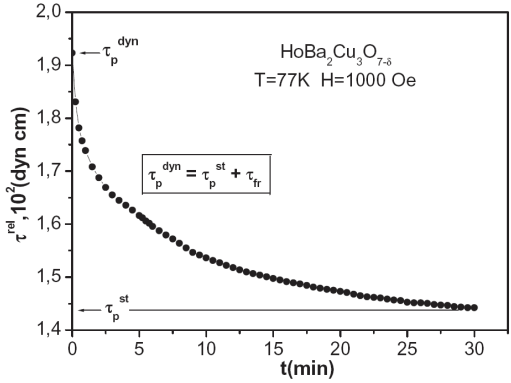

Fig.3. Dependence of momentum $\tau^{rel}$ on time $t$ after the stopping of rotating head for $HoBa_2Cu_3O_{7-\delta}$ sample at T=77K and H=1000 Oe.

If during the relaxation after rotation of sample one applies the pulsed magnetic field in parallel to the outer magnetic field $\vec{H}$, then additional vortices, created as result of magnetic pulse, influence the structure already existing in the sample as "the vortex fan" what could result in the decrease of the angle of its unfolding or to its folding. The letter in its turn, would cause the additional change in the relaxation process taking place in the sample, and, correspondingly, results in the stepwise decrease of moment related with viscous forces $\tau_{fr}$.

But the change of relaxation process character and, correspondingly, the stepwise decrease of moment could happen if the duration of magnetic pulse is larger as compared with the time necessary for creation of a new vortex structure, which will influence the superconducting sample relaxing in magnetic field. If it is the case, then at the small durations of magnetic pulses the relaxation curve, presented in Fig.3, doesn't change, but when this duration becomes the order of a time for penetration of vortices into the sample and the creation of vortex structure, then the aforementioned change of relaxation processes could principally appear. Namely, this situation when the duration $\Delta x$ of magnetic pulses was larger then the time for Abrikosov vortex lattice creation $\Delta x_c$, have been described by us our previous work [19], when it was shown that the influence of one magnetic pulse $\Delta h\approx 400$ Oe ($\Delta h//H$) with duration 30μсек>$\Delta x_c$ was stepwisely decreased the $\tau^{rel}$ moment and the relaxation process continued with the reduction of $\tau^{rel}$ on a level as far as a new magnetic pulse similar the first one is not applied.

In the presented work it was studied the influence of different duration and amplitude pulses on relaxation processes in vortex matter. The results shown in Fig.4 on action of single pulses of different durations on relaxation processes in vortex matter and, consequently, on mechanical moment $\tau^{rel}$ revealed that at small pulses durations up to 15 μs the $\tau^{rel}$ doesn't change, but at duration of applied pulse >15 μs it is observed the stepwise change $\tau^{rel}$, what speaks on the existence of the $\Delta x_c$ threshold.

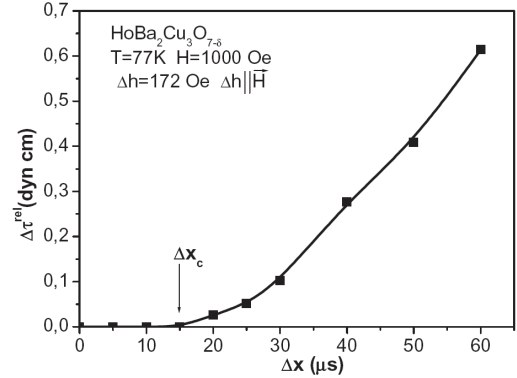

Fig.4. Dependence of momentum $\Delta\tau^{rel}$ on the duration $\Delta x$ of magnetic field single pulse $\Delta h$=172 Oe applied in parallel to the main magnetic field H=1000 Oe at T=77K for $HoBa_2Cu_3O_{7-\delta}$ sample.

This way one could say that the Abrikosov vortex lattice creation time in high-temperature isotropic superconductor of $HoBa_2Cu_3O_{7-\delta}$ makes value on the order of 20μs. This value approximately on the order of value higher then time for the single-vortex creation for the first time measured by G. Boato, G.Gallinaro and C. Rizzuto [28], who showed that this time is less than $10^{-5}$ sec.

In work [19] it was also shown that continuous action of aforementioned pulses with the train frequency equal to 2,5 $s^{-1}$ more sharply reveals their influence on relaxation processes in vortex matter and in these conditions the processes of penetration of vortices into superconductors bulk are made more sharply expressed. In fig.5 it is presented the clear picture of magnetic pulses continuous action with $\Delta h$=172 Oe ($\Delta h//H$), and the duration 20μsec, what is larger than the $\Delta x_c$ with the train frequency $\nu$ = 2,5 $s^{-1}$. As it is seen from picture the pulses of 5, 10 and 15 μsec durations doesn't change $\tau^{rel}=f(t)$ which is observed at absence of magnetic pulses.

The results presented in Fig.5 show that at durations of pulses in 20μsec, 30μsec and 40μsec the Abrikosov vortices penetrate into the superconductor. This way the threshold value on the magnetic pulses duration observed at the action of single pulses (Fig.4) coinside with the threshold observed when their repetition frequency is $\nu$ = 2.5 $s^{-1}$.

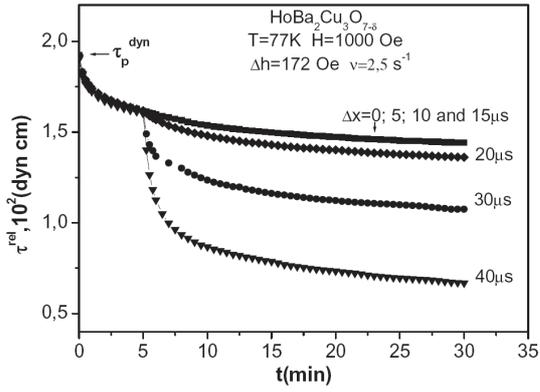

Fig.5. Dependence of momentum $\tau^{rel}$ on time $t$ after the stopping of rotating head with the influence since t=5 min on the relaxation process of HoBa$_2$Cu$_3$O$_{7-\delta}$ sample of the continuous magnetic field pulses   h=172 Oe  with  ν=2,5 s$^{-1}$  frequency and different durations     x=5; 10; 15; 20; 30; and 40 μs. Pulsed magnetic field was parallel to the main magnetic field  H=1000 Oe  at  T=77K.

In Fig.6 it is presented the curve of $\tau^{rel}$=f(t) dependence on time at the influence of magnetic pulses Δh=172 (Δh//H) the duration of which is below the time of Abrikosov vortex system creation Δx=5μs<Δx$_c$ (Δx$_c$≥15μs for the investigated HoBa$_2$Cu$_3$O$_{7-\delta}$). As it is seen from the picture when Δx<Δx$_c$, the relaxation curve doesn't change in spite the increase of the repetition frequency of magnetic pulses ν from 2.5 up to 500 s$^{-1}$. As soon as the duration of pulses exceeds the critical value and becomes  Δx=30μs, the relaxation curve undergoes the essential (stepwise) change. For example in Fig.6 it is presented measurement for ν=5s$^{-1}$  и ν= 500s$^{-1}$.

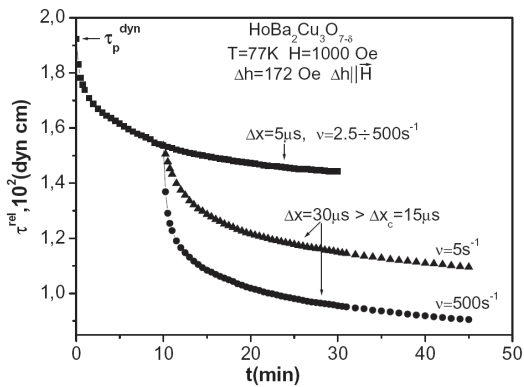

Fig.6. Dependence of momentum $\tau^{rel}$ on the time $t$ after the stopping of rotating head with the influence since  t=10 min  on the HoBa$_2$Cu$_3$O$_{7-\delta}$ sample relaxation process of  the continuous pulses magnetic field with frequency ν=2,5 ÷500s$^{-1}$ at Δx=5μs< Δx$_c$ , and also at Δx= 30μs > Δx$_c$. The pulsed magnetic field Δh=172 Oe was parallel to the main magnetic field  H=1000 Oe  at T=77K.

And finally, we have observed the threshold on the value of applied pulses. In Fig.7 it is shown that in spite the fact that we applied magnetic pulses of the  large duration 300μs>>Δx$_c$, much longer as compared with the time of Abrikosov vortex creation at small amplitudes of pulsed field Δh ~7, 11, 14 Oe  $\tau^{rel}$=f(t) doesn't change.

The stepwise change of the relaxing moment $\tau^{rel}$ is revealed only at Δh ~18 Oe  and higher.

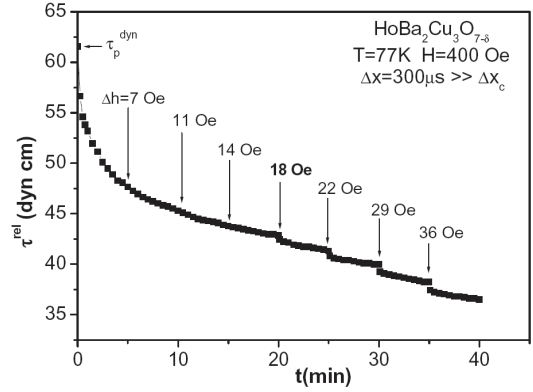

Fig.6. Dependence of momentum $\tau^{rel}$ on the time $t$ after the stopping of rotating head with the application after 5 minutes on the HoBa$_2$Cu$_3$O$_{7-\delta}$ sample relaxation process of  the single magnetic field pulses Δh=(7÷36) Oe  with duration Δx= 300μs >>Δx$_c$. The pulsed magnetic field was parallel to the main magnetic field H=400 Oe at T=77K.

The further investigations of relaxation phenomena are anticipated for anisotropic high-temperature superconductors among them in strongly anisotropic high-temperature superconductors of  Bi-Pb-Sr-Ca-Cu-O system.

## 4. Conclusion

The simple mechanical method of Abrikosov vortex stimulated dynamics investigations it was applied for the study of pulsed magnetic fields influence on relaxation phenomena in vortex matter of high- temperature superconductors. It was observed the change of relaxation processes in vortex matter as a result of pulsed magnetic field influence on it.

The study of influence of different duration and amplitude pulsed magnetic fields influence was revealed the existence of threshold phenomena. A small duration pulse doesn't change the course of relaxation processes in vortex matter of isotropic high- temperature superconductor HoBa$_2$Cu$_3$O$_{7-\delta}$. When the duration of pulses exceeds some critical value (threshold), then their influence change the course of relaxation processes. The latter is revealed in a stepwise decrease of relaxing mechanical momentum $\tau^{rel}$, apparently, related with a sharp change of pinning and the rearrange of vortex system of superconducting sample as a result of penetration into its bulk of a new portion of vortices at application of pulsed field on the outer magnetic field creating the main vortex structure in the investigated HoBa$_2$Cu$_3$O$_{7-\delta}$ sample. A new portion of vortices "shakes" the vortex lattice existing in a sample causing the detachment of vortices from a weak pinning centers what, apparently, is the reason for the stepwise decrease of mechanical momentum $\tau^{rel}$ .

All these made it possible to define the Abrikosov vortex lattice creation time in $HoBa_2Cu_3O_{7-\delta}$ which turned out to be on the order of value higher as compared with the time of single- vortex creation observed in type II superconductors.

**Acknowledgements**

The work was supported by the grants of International Science and Technology Center (ISTC) G-389 and G-593.